\documentclass{bioauto}
 \usepackage[abs]{overpic}
 \usepackage[numbers]{natbib}
\renewcommand{\vec}[1]{{\mathbf{#1}}}
\DeclareMathOperator{\sgn}{sgn}

\usepackage{soul}
\usepackage{xcolor}
\renewcommand{\hl}[1]{#1}

\newcommand{\rev}[1]{#1}
\usepackage[normalem]{ulem}

\BIOjournalYear{2021}
\BIOjournalVolume{XX}
\BIOjournalNumber{X}
\BIOjournalPages{XXX-XXX}

\BIOtitle{Estimation of Parameters for an Archetypal Model of Cardiomyocyte Membrane Potentials}

\BIOauthors{Muhamad H.N. Aziz, Radostin D. Simitev}

\BIOaddress{Institute of Mathematical Sciences \\
Universiti Malaya\\
50603 K. Lumpur\\
Malaysia \\
E-mail: \BIOauthorEmail{hifz$\_$din@um.edu.my},%
School of Mathematics and Statistics \\
University of Glasgow \\
Glasgow G12 8QQ\\
United Kingdom\\
E-mail: \BIOauthorEmail{Radostin.Simitev@glasgow.ac.uk}\\
\href{https://orcid.org/0000-0002-2207-5789}{orcid.org/0000-0002-2207-5789}
}

\BIOeditor{filled by editor}{filled by editor}{filled by editor}

\BIOabstract{
Contemporary realistic mathematical models of single-cell cardiac electrical
excitation are immensely detailed. Model complexity leads to parameter
uncertainty, high computational cost and barriers to mechanistic
understanding. There is a need for reduced models that are
conceptually and mathematically simple but physiologically accurate.
To this end, we consider an archetypal model of single-cell cardiac excitation that
replicates the phase-space geometry of detailed cardiac models, but at
the same time has a simple piecewise-linear form and a relatively low-dimensional
configuration space.
In order to make this archetypal model practically applicable, we
develop and report a robust method for estimation of its parameter values from
the morphology of single-stimulus action potentials derived from detailed ionic current
models and from experimental myocyte measurements.
The procedure is applied to five significant test cases and an excellent
agreement with target biomarkers is achieved.
Action potential duration restitution curves are also computed and
compared to those of the target test models and data, demonstrating
conservation of dynamical pacing behaviour by the fine-tuned archetypal model.
An archetypal model that accurately reproduces a variety of wet-lab
and synthetic electrophysiology data offers a number of specific
advantages such as computational efficiency, as also demonstrated in the
study. Open-source numerical code of the models and methods used is
provided.
}

\BIOkeywords{Mathematical models, Cardiac action potential, Electrophysiology, Parameter estimation. }

\DeclareFontEncoding{TS1}{}{}
\DeclareFontSubstitution{TS1}{cmr}{m}{n}
\DeclareTextSymbol{\tcrp}{TS1}{'251}
\DeclareTextSymbolDefault{\tcrp}{TS1}
\newcommand{\licen}{
\begin{tabular}{p{0.17\textwidth} p{0.79\textwidth}}
\raisebox{-30pt}{\includegraphics[scale=0.5]{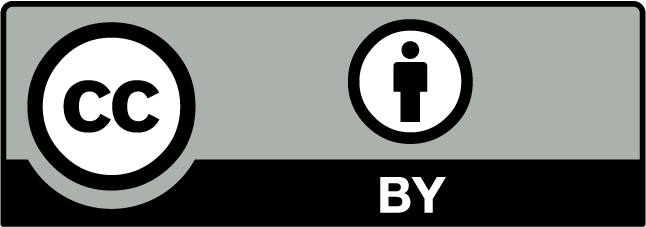}}	& {\fontsize{10}{10}\selectfont
	\tcrp ~2018 by the authors. Licensee Institute of Biophysics and Biomedical Engineering, Bulgarian Academy of Sciences. This article is an open access article distributed under the terms and conditions of the Creative Commons Attribution (CC BY) license \phantom{httppppp} \urlstyle{same} (\url{http://creativecommons.org/licenses/by/4.0/})}.
\end{tabular}
}

\begin{document}

\biotitle

\section{Introduction}

\looseness=-1
Models of the action potential of cardiac cells are routinely used to
interpret and integrate experimental findings, extrapolate animal
data to human system context, and test novel hypotheses
\cite{Heijman2015,Trayanova2021}. It is frequently proposed that 
these models will soon make it possible to  devise patient-specific
precision therapies, and accelerate cardiac drug discovery and
development \cite{Amanfu2011,Davies2016,Trayanova2020}.
Since the pioneering work of Noble \cite{Noble1962}, over 150 models
have been published  with the aim to capture in detail
the electrophysiology of a wide variety of cardiac cell types under a broad
range of experimental, physiological and pathological conditions \cite{Fenton2008}. Contemporary detailed models largely follow the
Hodgkin-Huxley paradigm but have grown to a staggering
complexity \cite{Sigg2010}. For example, a recent model of the human ventricular
action potential \cite{OHara2011} consists of 49 ordinary differential 
equations and includes 206 model parameters. Because models are
typically developed by extending and re-using components from
previous models, as advocated by large international initiatives like the
Physiome Project \cite{Bassingthwaighte2000} and the CellML Project
\cite{Miller2010}, many of these parameters and equations are poorly 
constrained and in many cases redundant. 
For instance, the meta-analysis of \cite{Niederer2009} demonstrates that the
modern human ventricular models \cite{tenTusscher2004,Iyer2004}
include parameters that have been inherited from studies in at least 9
different species over a range of 6 different temperatures; in other
words, it is questionable whether they represent any human ventricular
myocyte.
While intense research effort is expended to estimate parameter uncertainty
\cite{Clayton2020}, calibrate models to identifiable and reliable
experimental protocols \cite{Whittaker2020,Clerx2019}, increase
reproducibility \cite{Cooper2016} and build trustworthy models
\cite{Johnstone2016}, detailed cardiac cell models remain
difficult to benchmark and to adapt to situations to which they have
not been fitted \cite{Wilhelms-2013} and are computationally expensive
especially in tissue-scale simulations \cite{Clayton-2011}. Most importantly,
detailed cardiac models are becoming increasingly difficult for causal
inference \cite{Biktashev-2008}. 

\looseness=-1
Thus, there is a certain need for simplified mathematical models
that are accurate and flexible enough, computationally affordable, 
amenable to mathematical analysis and to mechanistic 
understanding. Starting with the early work of van der Pol \cite{vanderPol1928}
a number of conceptual models have been proposed to address this need,
e.g.~\cite{FitzHugh1961,Nagumo1962,Aliev1996,Fenton-Karma-1998,Mitchell-Schaeffer-2003},
and most of them have become popular and frequently used in the place of
detailed ones. However, these conceptual models rely mostly on ad hoc 
assumptions and generally have a FitzHugh-Nagumo structure that leads
to certain shortcomings \cite{Biktashev-2008}.
In contrast, in \cite{Biktasheva2006,Biktashev-2008} we  developed an asymptotic
method that allows for a systematic and 
controlled reduction of arbitrary detailed cardiac ion current
models.
The method preserves the phase-space geometry of detailed
models, different from the FitzHugh-Nagumo one, and reveals qualitatively new
features of topological nature \cite{Simitev2011}.
Following this approach in \cite{Biktashev-2008} we reduced Noble's model
of purkinje fibre electrophysiology \cite{Noble1962}. We obtained a
mathematically simple conceptual model that consists of three piece-wise linear
differential equations and contains only 13 intrinsic model
parameters and so it is rather inexpensive to integrate numerically. 
Further, the model admits closed-form analytic solutions when
spacially-clamped \cite{Biktashev-2008}, and closed-form travelling
wave solutions when spacially-extended \cite{Simitev2011}. These exact solutions aid
mechanistic understanding, extensive exploration of parameter space
as well as benchmarking of numerical codes. More importantly the  model is
archetypal in the sense that it has the   generic asymptotic structure
of modern detailed cardiac ionic models and it is thus capable of
reproducing slow repolarization, slow sub-threshold response, fast
accommodation, front dissipation, variable peak voltage and other
features of cardiac excitability crucial for understanding and
controlling arrhythmogenesis \cite{Biktashev-2008} where most other
ad hoc conceptual equations often fail.

\hl{In order to be practically useful beyond its utility as a conceptual
tool, the parameter values  of the archetypal model} 
\cite{Biktashev-2008} \hl{must be determined so that it replicates the
behaviour of state-of-the art models of ventricular and atrial
excitation and captures experimental 
measurements quantitatively. This is the goal of the present work.}  To this end we
describe in the following the implementation of a standard parameter
estimation procedure and use it
to fit the archetypal model to a typical mammalian ventricular model
\cite{Luo-Rudy-1991}, a typical human atrial model 
\cite{Courtemanche-1998}, as well as to experimental data for rabbit
ventricular myocytes available from the literature
\citep{McIntosh-2000}. In addition, we provide an open-source
numerical code permanently available at \cite{Aziz2021} that
can be used by the reader to apply the methodology to other detailed
models and data of their own interest.
Further, fitting data and detailed models to a common set of
equations gives the opportunity to compare and contrast such
models directly, which is otherwise impossible due to their different
mathematical structure and components.
\rev{The scale of the computational effort required to perform
  large-scale tissue and whole heart simulations is immense and
  simulations in real-time are beyond current computational
  capability.  A number of  strategies for reducing calculation time
  are used at present including  code parallelisation, lookup tables,
  exponential solutions for gating variables, operator splitting,
  adaptive time and space stepping, using graphics processing units
  and using simplified models \cite{Clayton-2011}}. \rev{In this
connection,} from a software engineering viewpoint it could be very
beneficial to be able to replace the variety of different cell
models, e.g.~ventricular, atrial, sinoatrial, involved in large-scale
tissue and whole-heart simulations, by a single model, \rev{as
  proposed here}, with different parameter values at different spacial
positions.

\section{Model formulation and methods of parameter estimation}

\subsection{The archetypal cardiac cell model}
\label{analytics}

\looseness=-1
Cardiac cell membranes are composed of a biphospholipid layer,
impermeable to charged particles and maintaining a non-zero
equilibrium voltage potential across the membrane \cite{Sigg2010}. The layer is
protruded by voltage-gated ion channels -- large proteins that open
and close depending on the instantaneous value of the voltage and
allow in/outflux of ion currents. When a cell is ``excited'', these currents cause the
formation of a large transmembrane voltage excursion known as an action potential.
The action potential propagates within the myocardium and signals
cardiac cell contraction thus controlling the heartbeat
-- the main function of a living heart.   
To a first approximation the physiology of cell membranes can be modelled as an
electrical circuit consisting of a capacitor $C_m$ representing the biphospholipid
layer, and an active resistor supporting ionic currents
$I_\text{ion}(E)$ representing ionic channels, that are connected in
parallel, giving rise to the ordinary differential equation $C_m
\dot{E}=I_\text{ion}(E)$ for the transmembrane voltage potential
$E(t)$ \cite{Fenton2008}.  
The models of the ionic current $I_\text{ion}(E)$ encapsulate the
electrophysiological properties of the cardiac membrane mentioned
above and over the
last 70 years have grown to a staggering complexity as discussed in
the Introduction.

\looseness=-1
Here, we consider the following archetypal model for the action potential of
a single cardiac cell
\begin{subequations} \label{model}
\begin{align}
&\frac{d}{dt} E = \frac{1}{\epsilon_1
  \epsilon_2}G_\text{Na}\,\Big(E_\text{Na}-E\Big)
  \theta(E-E_{\ast})\,h  + \frac{1}{\epsilon_2} \Big(g_{2}(E)\, n +
  G(E)\Big), 
  \label{modelEqn1}   \\
&  \frac{d}{dt}h =  \frac{1}{\epsilon_1 \epsilon_2}f_h\,\Big(\theta(E_{\dagger}-E)-h \Big) , \label{modelEqn2} \\
&  \frac{d}{dt}n = F_n(E)\, \Big(\theta(E-E_{\dagger}) -
  n\Big).  \label{modelEqn3}
\end{align}
The model takes the form of a set of piecewise-linear ordinary
differential equations for the evolution in time $t$ of three state
variables -- the voltage  $E$, and the gating
variables $h$ and $n$ describing the inactivation of a fast inward current
and the activation of the time-dependent channel of a slow outward
currents, respectively. Here $\theta(.)$ is the Heaviside step function and 
\begin{align}
\label{g2eq}
& g_2(E) = g_{21} \theta(E_{\dagger} - E) + g_{22}
\theta(E-E_{\dagger}), ~~~~~
F_n(E) = f_n\,\Big(r \theta(E_{\dagger}-E) + \theta(E-E_{\dagger})\Big), \\
\label{Geq}
&  G(E) =
  \begin{cases}
        k_1\big(E_1 - E\big), & E \in (-\infty,E_{\dagger}), \\
        k_2\big(E - E_2\big), &  E \in [E_{\dagger},E_{\ast}), \\
  	    k_3\big(E_3 - E\big), &  E \in [E_{\ast},+\infty),
  \end{cases}\\
  \label{E2eq}
& E_2 = \Big(k_1/k_2 + 1\Big)\,E_{\dagger}-E_1 k_1/k_2, ~~~~~
E_3 = \Big(k_2/k_3 + 1\Big)\,E_{\ast} - E_2 k_2/k_3.
\end{align} 
\rev{The constants \hl{$f_h$, $f_n$} are time scales of approach
  of channel gating variables $h$ and $n$ to their respective steady
  states, with the parameter \hl{$r$} further modulating $f_n$. The steady
  states of $h$ and $n$ are assumed to have a  ``perfect switch''  behaviour and the
  parameter \hl{$E_{\dagger}$} is the voltage value at 
  which switching occurs. An additional unnamed ultra-fast gating
  variable that switches instantaneously at \hl{$E_\ast$} is implicitly
  included in the model by the factor $\theta(E-E_{\ast})$.
  This is a conventional Hodgkin-Huxley description of channel gating
  kinetics \cite{Keener-Sneyd-2009} and the perfect switching assumption is
  often used in other simplified models \cite{Fenton-Karma-1998,Hinch2002}.  
  In addition to the time-dependent channel, the slow outward current
  has an instantaneous voltage-dependent channel modelled by the
  three-branch N-shaped piece-wise linear function \hl{$G(E)$} with branch
  slopes  \hl{$k_1$, $k_2$, $k_3$} and intercepts \hl{$E_1$, $E_2$} and \hl{$E_3$} of which only $E_1$ is an
  independent parameter because the branches are assumed to intersect
  at $E_{\ast}$ and $E_{\dagger}$. The constants \hl{$G_\text{Na}$,
  $g_{21}$, $g_{22}$} represent maximal ion channel conductances of the
  fast inward and the time-dependent slow outward currents,
  respectively, and \hl{$E_\text{Na}$} is the peak voltage.  This structure of the
  equations bears some resemblance to Nobel's model of purkinje fiber
  cells \cite{Noble1962} from which it derives, as discussed further
  bellow. However, to permit a general interpretation, the   currents
  in our model are not explicitly identified with the fast sodium and
  the slow potassium and leakage currents of the Nobel model.}  

Equations \eqref{modelEqn1} to \eqref{modelEqn3} are integrated in
time starting from the initial conditions  
\begin{gather}
  \label{stimvolt1}
E(0)=E_\text{stim}>E_\ast,~~~~h(0)=1,~~~~n(0)=0,
\end{gather}
and then advancing the solution via a sequence of initial value
problems on time intervals $t\in\big(kB,(k+1)B\big]$, $k=1,2,\dots$
with duration \hl{$B$} (basic cycle length, \rev{BCL}), and with initial conditions
\begin{gather}
    \label{stimvolt2}
  E(kB)=E_\text{stim},~~~~h(kB)=h\big((k-1)B\big),~~~n(kB)=n\big((k-1)B\big).
\end{gather}
\end{subequations}

\looseness=-1
The archetypal model \eqref{model} was first introduced in
\cite{Biktashev-2008} as an asymptotic embedding of the original Noble
purkinje fiber equations \cite{Noble1962} using a set of verifiable transformations
with simplification errors that can be measured and controlled
accurately. For full details we refer to 
\cite{Biktashev-2008}, and here we only note briefly that the asymptotic embedding procedure
takes into account the following empirical properties that the
original Noble model has in common with the vast 
majority of other detailed ionic current models: 
(a) the large differences in the time-scales for evolution of state
variables, (b) the large maximal  value of the sodium current
$I_\text{Na}$ compared with other currents and (c) the
quasi-stationary permeability of the $I_\text{Na}$ ionic gates in
certain potential ranges. Equations \eqref{model} differ from the 
system introduced in \cite{Biktashev-2008} only in that $n$ instead of $n^4$ is
used in equation \eqref{modelEqn1}, with the goal to obtain a fully
linear system with even simpler closed-form solutions and because
parameter values will be adjusted anyway as discussed further
below. The 
archetypal model \eqref{model} has already been used to derive
asymptotic expressions for the conduction velocity restitution in
cardiac tissues \cite{Simitev2011}, to understand the formation of
excitation waves \cite{Bezekci2015}, and most recently its fast-time
subsystem was employed to elucidate the conditions for arrhythmogenesis
and refractoriness in atrial tissue with myocyte-fibroblast coupling
\cite{Mortensen2021,Mortensen2021b}.

Problem \eqref{model} has 17 free parameters. Four of them
are not intrinsic: \hl{$\epsilon_1, \epsilon_2, E_\text{stim}$} and
$B$. The positive constants \hl{$\epsilon_1, \epsilon_2\in[0,1]$}  
are asymptotic parameters embedded in the model to enable formal asymptotic
analysis but both will be kept fixed to unity in the present
study. The stimulus voltage $E_\text{stim}$ and the basic cycle length
$B$ are typically specified as a part of an external periodic
cell stimulation protocol, also cf.~equation \eqref{Istimeq} below. The remaining 13 parameters are intrinsic
to equations \eqref{modelEqn1} to \eqref{E2eq} and we represent them
as the components of a column vector 
\begin{gather}
\label{peq}  
\vec{p}= [k_1, k_2, k_3, E_1, E_{Na}, E_\dagger, E_*, f_h, f_n, r,
  G_{Na}, g_{21}, g_{22}]^T.
\end{gather}
The objective of the study is to find appropriate values for these
parameters as discussed next.

\subsection{Target models and experimental data}

We seek to estimate the values of the protocol-independent parameters
\eqref{peq} of the archetypal model \eqref{model} so that the model
outputs reproduce the behaviour and the biomarkers of target detailed
ionic models or experimental measurements. Here, as target
models/data we consider the models of Noble \cite{Noble1962}, Luo-Rudy
\cite{Luo-Rudy-1991}, and Courtemanche \cite{Courtemanche-1998}, as
well as the measurements of \citep{McIntosh-2000} for rabbit
ventricular myocytes. 

The Noble model \cite{Noble1962} describes the action potential of
Purkinje fibre cells. It incorporates a sodium current and two
different types of potassium current. The model is based on
the Hodgkin-Huxley formulation adjusted to the action potential of Purkinje cells, which is
significantly different from that of the squid giant axon in terms of
plateau duration. This model is the first ever mathematical model of the
action potential of cardiac myocytes and is the ancestor of most
current detailed ionic models and the basis of the archetypal model
\eqref{model}. 

The Luo-Rudy (LR) model \cite{Luo-Rudy-1991} captures the single-cell
ventricular action potential of guinea pigs. It includes six ionic
currents (sodium, slow inward, time-dependent, time independent and
plateau potassium currents and a background current) controlled by
seven gate variables and a description of the intracellular calcium
concentration.  

\looseness=-1
The Courtemanche et al.~(CRN) model \cite{Courtemanche-1998} describes the
action potential of human atrial myocytes. It has 13 ionic
currents including formulations of K+, Na+ and Ca2+ currents
and representations of pump, exchange and background currents.
The model is capable of responding to rate changes, calcium channel
inhibition and sodium-calcium pump exchanger blockade.

Machine readable implementations of three mathematical models
are available from the CellML model repository \cite{Miller2010}, and are also included
with our code \cite{Aziz2021}.
The models are supplemented by stimuli currents
$I_\text{stim}(t)$ that 
take the form of periodic trains of rectangular impulses with
amplitude $I_s$, duration $t_s$, and period (basic cycle length) $B$,
\begin{gather}
  I_\text{stim}(t)=I_s\, \left[1+ \sgn \left(\sin\frac{\pi t}{B}\right) \sgn \left(\sin\frac{\pi(t-B-2 t_s)}
{B}\right)\right].
\label{Istimeq}
\end{gather}
\looseness=-1
The inclusion of these currents is known as ``stimulation by
current''. The currents are used to excite action potentials in the same
way as in an experiment and provide an equivalent alternative to the 
``stimulation by voltage'' given by equations \eqref{stimvolt1} and
\eqref{stimvolt2} for the archetypal model.

The experimental recordings of \cite{McIntosh-2000} consist of
measurements of action potential and intracellular calcium transient
characteristics in isolated myocytes from sub-epicardial,
mid-myocardial and sub-endocardial regions of the rabbit left
ventricles. These measurements were recorded under both healthy and
heart failure conditions. Results showed that in the heart failure
group, AP duration and  calcium transient duration were prolonged in
both sub-epicardial and mid-myocardial cells. These changes were
significant at lower stimulus frequencies but the relative effect
diminished at higher frequencies. Below we will only consider the
measurements of mid-myocardium cells in both healthy and failing
myocytes.

\subsection{Numerical solution and biomarkers}

\looseness=-1
The archetypal model \eqref{model} and each of the three target models
are integrated in time with a relative tolerance of $10^{-6}$  employing an
adaptive-step, adaptive-order method for systems of stiff ordinary
differential equations based on the numerical differentiation formulas
of \cite{Shampine1997} as implemented in Matlab(TM) \cite{MATLAB} functions
\texttt{ode15s} and \texttt{ode23}. The resting states of the three target models were
used as their respective initial conditions. While a closed-form
analytic solution of the archetypal models is available  
\cite{Biktashev-2008}, a numerical solution is used here for
consistency with the target models. 

The following biomarkers are computed for each model. 
{
    \setlength{\parskip}{-8pt}
    \setlength{\parsep}{-8pt}
    \begin{enumerate}
    \setlength{\itemsep}{0pt}
    \setlength{\parskip}{0pt}
    \setlength{\parsep}{0pt}  
\item Discretized voltage trace during the $k$-th basic cycle
  length period $t\in\big(kB,(k+1)B\big]$. More precisely, for the archetypal model this takes the form
  of a set of ordered pairs $\big\{\big(t_{ki}, E_{ki}\big),~ i=0,\dots,
  M\big\}$, consisting of discrete values of time $t_{ki}=kB+i\,\Delta t$
  and discrete values of the voltage $E_{ki}=E(t_{ki})$ and $M=B/\Delta
  t$. A step of $\Delta t=0.1$ ms is used which is sufficient to
  resolve the traces.  
\item Action potential duration at 90\% of the voltage peak amplitude
  in the $k$-th basic cycle length period
  $\mathcal{I}=\big(kB,(k+1)B\big]$, defined as the solution $A_k$ of the equation
  \begin{gather}
    E(A_k+kB)= 0.9 \Big(\max_{t\in\mathcal{I}} E(t) - \min_{t\in\mathcal{I}}
      E(t)\Big),
  \end{gather}
  that satisfies the condition $\dot{E}(A_k+kB)<0$. We refer to this
  biomarker as $\text{APD}_\text{90}$, an abbreviation often used in
  the experimental and the physiological literature.
\end{enumerate}
\looseness=-1
The same type of biomarkers are also available from the experimental
measurements of McIntosh et al \cite{McIntosh-2000}. Their values were 
extracted from the published manuscript  using the
online tool WebPlotDigitizer \cite{Rohatgi2020}.
}

\subsection{Parameter estimation}

Having defined appropriate biomarkers, we now compare the
archetypal model \eqref{model} to each of the target models and the
experimental data using a residual (also known in the literature as ``error'' or
``cost'' or ``objective'')
function of the form
\begin{gather}
\label{residual}
R(\vec{p}) = \frac{1}{2}\left(\left|\frac{A(\vec{p}) -
  \mathcal{A}}{\mathcal{A}}\right|+ \frac{1}{M+1}\sum_{i=0}^M  \left|\frac{E_i(\vec{p})
  - \mathcal{E}_i}{\max\limits_{j=0..M} \mathcal{E}_j-\min\limits_{j=0..M} \mathcal{E}_j}\right|\right).
\end{gather}
Here, $E_i$ are the discrete values of the voltage trace, $A$ is the
$\text{APD}_\text{90}$ of the archetypal model while the calligraphic
symbols $\mathcal{E}_i$ and $\mathcal{A}$ denote the corresponding
values for the target models and data.
In particular, the
biomarkers in the initial basic cycle period $k=0$ are used and, for
brevity, the subscript $k$ is omitted. This form of the residual measures
the discrepancy in the morphologies of the action potentials
between the archetypal and the target models/data. The first term of
the residual allows the fitting algorithm to assign extra weight to
matching the value of $\text{APD}_\text{90}$  when optimising action potential morphology.
We remark that, in general, for a complete comparison of the models, the
residual needs to include differences between all dynamical
variables of the models compared, including all gating variables and
ion concentrations. This however, is not possible because of the
difference in model formulations, in particular, because different
models include different ionic currents and have different dynamical
variables. Similarly, model quantities are not easily measured in
experiments. The voltage is often the only quantity in common between
different models and between models and data.

We now compute parameter values $\widetilde{\vec{p}}$ of the archetypal
model \eqref{model} such that the residual \eqref{residual} is mimimised, in symbolic form,
\begin{gather}
  \label{argmin}
\widetilde{\vec{p}}= \arg\min\limits_{\vec{p}\in \Omega}\, R(\vec{p}),
\end{gather}
where $\Omega\subset \mathbb{R}^{13}$ is the ball $|\vec{p}-\vec{p}_0| <
  \tau$ centred at the default values $\vec{p}_0$ of   the archetypal
  model parameters given in the second column of Table \ref{table1}
  and radius $\tau=|\vec{p}_0|$. 
More explicitly, for the parameter estimation function
$\arg\min(\cdot)$, we use a MATLAB implementation 
\citep{derrico2021} of the bounded gradient-free Nelder-Mead simplex
method \citep{Nelder-Mead-1965,Lagarias1998} for minimisation of
real-valued multivariate functions.

We provide an open-source numerical code including the models and
methods described in this section. The code is permanently available at \cite{Aziz2021}
and can be used by the readers to reproduce the results described below
an/or to apply the methodology to other detailed models and data of their own interest.

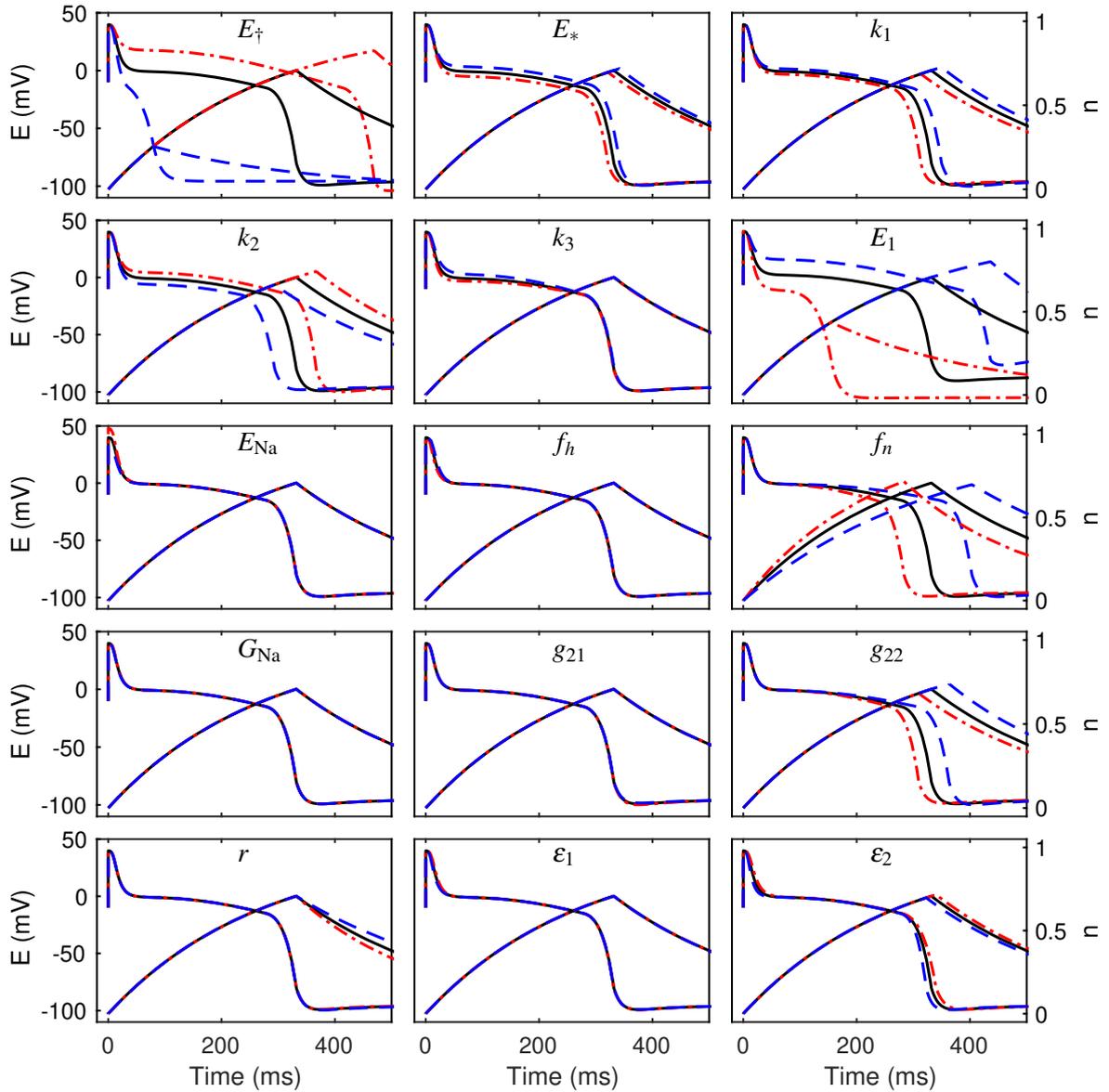
\begin{figure}[t!]
\centering
\begin{overpic}[width=\textwidth]{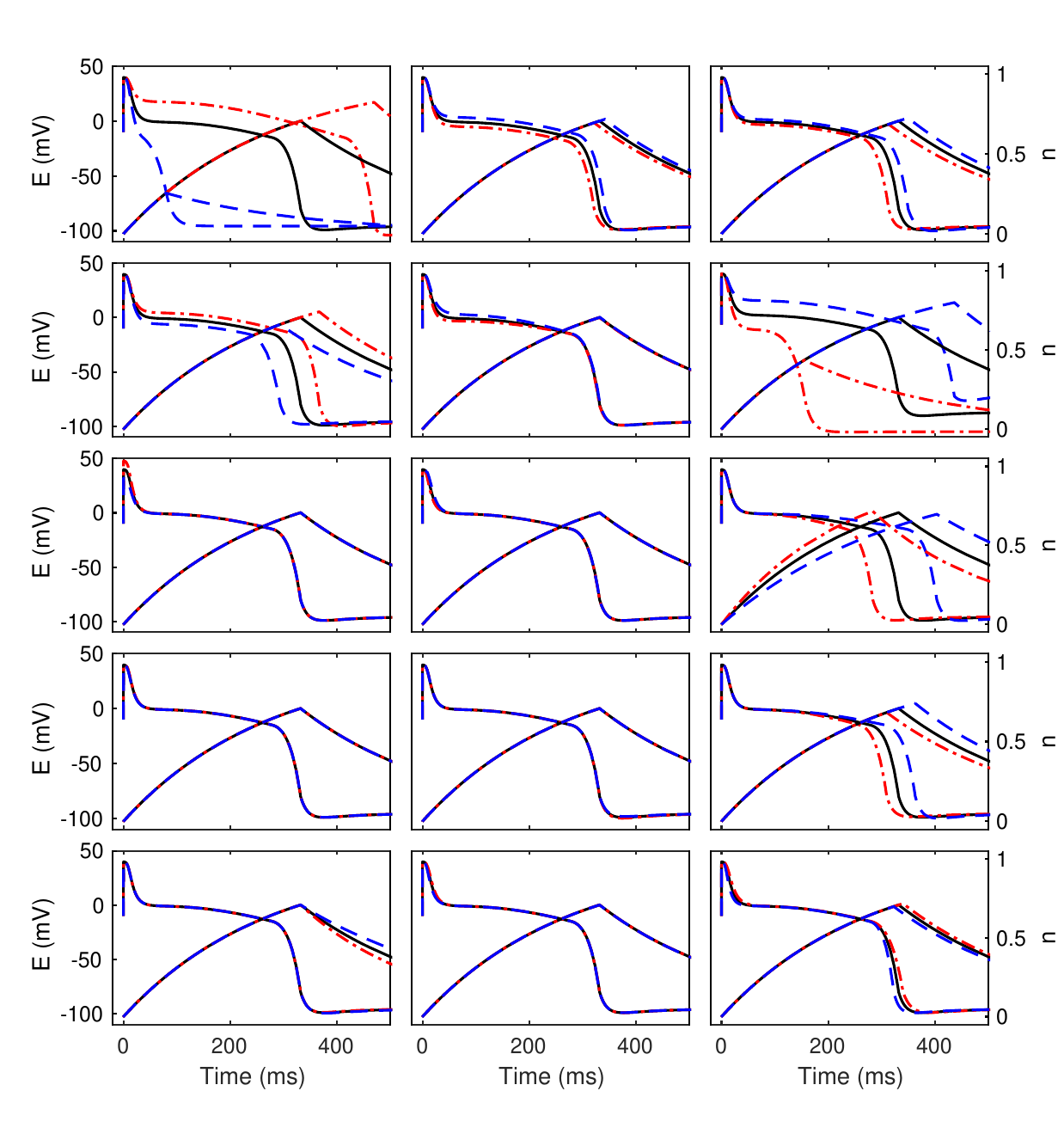} 
\put (105,447) {\small{$E_{\dagger}$}}
\put (232,447) {\small{$E_{\ast}$}}
\put (360,447) {\small{$k_{1}$}}
\put (105,362) {\small{$k_{2}$}}
\put (232,362) {\small{$k_{3}$}}
\put (360,362) {\small{$E_{1}$}}
\put (105,279) {\small{$E_\text{Na}$}}
\put (232,279) {\small{$f_{h}$}}
\put (360,279) {\small{$f_{n}$}}
\put (105,195) {\small{$G_\text{Na}$}}
\put (232,195) {\small{$g_{21}$}}
\put (360,195) {\small{$g_{22}$}}
\put (105,112) {\small{$r$}}
\put (232,112) {\small{$\epsilon_{1}$}}
\put (360,112) {\small{$\epsilon_{2}$}}
\end{overpic}
\vspace{-1.0cm}
\caption{Sensitivity of the profiles of the voltage $E$ and
  the slow gating variable $n$ of archetypal model
  \eqref{model} to the variation of a single parameter as denoted by
  the corresponding symbol in each panel. In all panels, the model outputs 
  generated from the default parameter values in Table \ref{table1} are shown by solid black
  line. The dashed blue line and dash-dotted red line are the model
  outputs obtained with $-20\%$ and $+20\%$ perturbation from the
  default, respectively.}
\label{fig1}
\end{figure}

\section{Results and analysis}

\subsection{Parameter sensitivity analysis}

Prior to estimating the parameter values \eqref{peq} of the archetypal
model \eqref{model}, we
perform a local sensitivity analysis in order to observe how each
parameter affects the action potential morphology and to establish whether
it is necessary to include all thirteen of them in the optimazation
search \eqref{argmin}. A formal local sensitivity analysis, see
e.g.~\cite{Snowden-2017}, involves
computing a sensitivity matrix $\vec{S}$ whose entries $S_{ij}(t)$
describe the normalised effect of perturbing the $j$-th parameter on  
the $i$-th state-variable, defined as 
\begin{gather}
  S_{ij} (t) \equiv  \frac{p_{j,0}}{x_i(t;\vec{p}_0)}  \left[\frac{\partial x_i(t;\vec{p})}{\partial
    p_j}\right]_{\vec{p}=\vec{p}_0} =  \left[\frac{\partial \log x_i(t;\vec{p})}{\partial
    \log p_j}\right]_{\vec{p}=\vec{p}_0}, ~~~~~ \vec{x}=[E,h,n]^T.
\end{gather}
\looseness=-1
However for clarity, instead of illustrating the elements of the
sensitivity matrix, we employ  
a simpler and more intuitive approach. We vary the value of
each of the thirteen model parameters by $\pm20$\% from their default
values listed in the second column of Table \ref{table1} at all other
parameter values fixed and observe how this variation affects the
traces of the voltage $E$ and the slow gating variable $n$.
Figure \ref{fig1} illustrates the results from this experiment and
also serves to illustrate clearly the effect each of the parameters
has on the action potential morphology.
For example, the peak membrane potential (PMP) is controlled by the
value of $E_\text{Na}$ while the resting membrane potential (RMP)  is
influenced by $E_1$ only. Some of the parameter values affect the action
potential morphology in a more complex way. For instance, $E_{\ast}$,
$E_{\dagger}$, $g_{22}$ and $f_n$ control repolarisation, but
$E_{\ast}$ and $E_{\dagger}$ also contribute to the duration of the
plateau.   
The fast gating variable $h$ exhibits negligible variation from its
quasi-stationary value $\bar{h}=1$ apart from two very short time
intervals during the front and the back of the action potential and
for this reason is not included in Figure \ref{fig1}.

\begin{table}[t]
\centering
 \begin{tabular}{|c| c c c c | c c|} 
\hline 
\multicolumn{7}{|c|}{(a) Parameter values} \\
\hline
 Params,  $p_i$	& Default 	& Noble \cite{Noble1962}	& LR \cite{Luo-Rudy-1991}	& CRN
 \cite{Courtemanche-1998} & Healthy \cite{McIntosh-2000} & Failing \cite{McIntosh-2000} \\ 
 \hline 
 
 $G_\text{Na}~[\text{ms}^{-1}]$ &100/3 &100/3 &100/3  &100/3 	&100/3 &100/3 	\\
 
 $E_\text{Na}$ [mV] &40.0 &40.0  &45.0 &24.3 & 41.5 & 41.5 \\
 
$E_{\dagger}$ [mV] &-80.0 &-80.0	&-75.0 &-60.0 &-80.0 &-78.0	\\ 
 
 $E_{\ast}$ [mV] &  -15.0 	& -10.0  	&11.5		&-9.0			&20.0 & 20.0		\\
 
 $k_1$ 		& 0.075 	& 0.04932  	&0.03602	&0.01702		& 0.0231  &  0.0173		\\
 
 $k_2$ 		& 0.04 	& 0.03033  	&0.00443	&0.007173			&   0.0057   & 0.0068	\\
 
 $k_3$ 		&  0.10 	& 0.08007  	&0.33946	&0.99977		&  0.0471   &   0.031		\\
 
 $E_1$ [mV] 		&  -93.333 	& -95.667 	&-84.333 	& -81.667		&-245/3 & -245/3		\\
 
 $f_n~[\text{ms}^{-1}]$ 		&  0.0037 	& 0.004471 	& 0.003781 	&0.00353		&   0.0043     &  0.0040	\\
 
 $f_h~[\text{ms}^{-1}]$ 		&  0.5 	& 0.5 		& 0.5 		&0.5				& 0.5 & 0.5	\\
 
 $g_{21} [\text{mVms}^{-1}]$ 		& -1.0 	& -0.28325 	&-0.14359 	& -0.02303		&  0.0409   &  0.0405		\\
 
 $g_{22} [\text{mVms}^{-1}]$ 		&  -9.0 	& -2.15744  	&-0.36512 	&-0.25370	& -1.097   & -0.9461 		\\
 
 $r$  			& 1.0 		& 0.7	 	&1.8		&2.8		& 0.4   & 0.6		\\  [1ex] 
 \hline
 \multicolumn{7}{|c|}{(b) \hl{Residual error of AP morphology} \eqref{residual}} \\
 \hline
$R(\vec{p})$ &-  &0.0067  &0.0250  &0.0235  &0.0155 &0.0118 \\ 
 \hline
\end{tabular}
 \caption{Estimates of the parameter values of the archetypal model
   \eqref{model} to
selected target models and data as described in text.
(a) Parameter values fitted by \eqref{argmin}. (b) The residual error
in action potential morphology \eqref{residual} between the archetypal
model and targeted models/data.  
The time is measured in ms, other units are given in the table if dimensional.}
\label{table1}
\end{table}

\subsection{Parameter estimation}

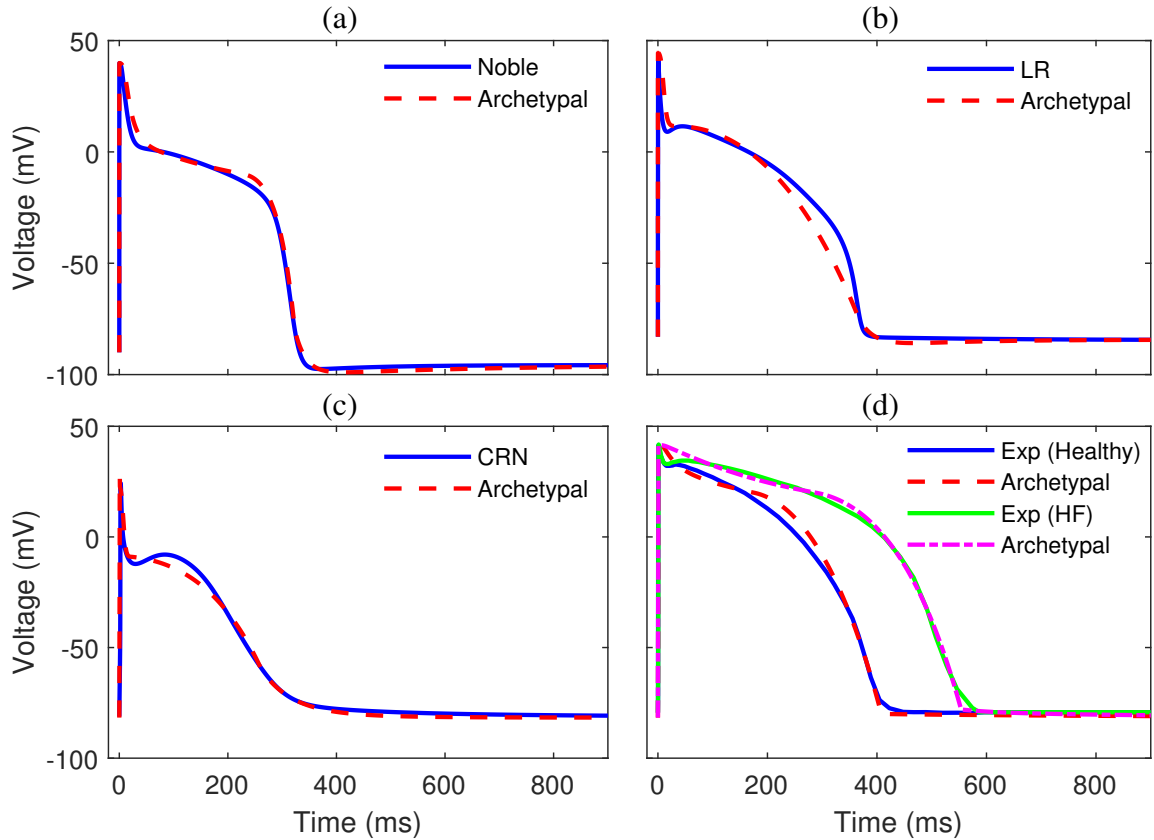
\begin{figure}[t]
\centering
\begin{overpic}
[width=\textwidth]{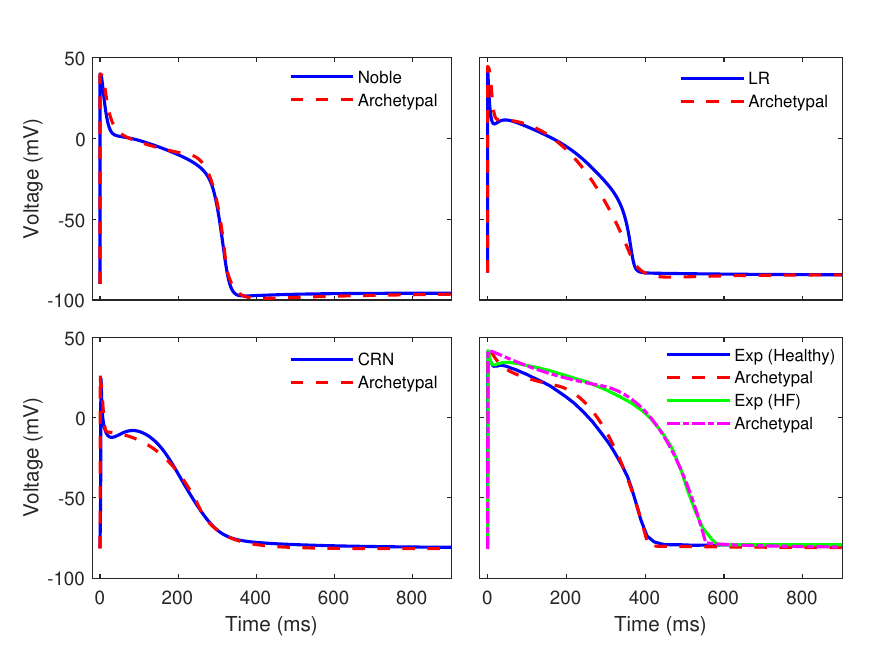} 
\put (128,316) {{(a)}}
\put (330,316) {{(b)}}
\put (128,170) {{(c)}}
\put (330,170) {{(d)}}
\end{overpic}
\caption{Agreement in action potential morphology between the
  archetypal model at parameter values listed in Table \ref{table1}
  (broken lines) and corresponding target models (panels (a)-(c),
  solid lines) and experimental data  (panel (d), solid lines). Model
  names are specified in the panel legends.
}
\label{fig2}
\end{figure}

Table \ref{table1} lists the results of applying the minimisation
procedure \eqref{argmin} to estimate the parameter values \eqref{peq} of the
archetypal model \eqref{model} so that it closely reproduces the
action potential morphology of the target models
\cite{Noble1962,Luo-Rudy-1991,Courtemanche-1998}  and the target data
\citep{McIntosh-2000}. The agreement obtained in action potential
morphology between the archetypal model and the targets is shown in
Figure \ref{fig2}. Target models and the archetypal 
model were stimulated at a basic cycle length $B=1500$ ms. 
The archetypal model is able to capture the action potential
morphologies of all target models and data used well.
The fits of the archetypal model to the CRN \cite{Courtemanche-1998}
and the LR \cite{Luo-Rudy-1991} models show the biggest 
discrepancy as seen in Table \ref{table1}(b) and Figures \ref{fig2}(b)
and (c).
The discrepancies are most significant in the neighbourhoods of the post
overshoot drop and plateau regions. In particular, the archetypal
model is somewhat inaccurate in capturing the deep notch produced by the CRN \cite{Courtemanche-1998} model.
This occurs because the archetypal model lacks transient outward
currents $\text{K}^+$ and $\text{Na}^+/\text{Ca}^{2+}$ exchanger
currents \citep{Santana-2010, Penaranda-2012} that act to form notch
phase  in the CRN \cite{Courtemanche-1998} model.
The fit of the archetypal model to the Noble 
model \cite{Noble1962} is rather satisfactory.
The peak voltage is controlled by the fast sodium current. The peak
voltage differs in different regions of the heart due to variation in
magnitude of the sodium current. In the archetypal model, the magnitude of
the fast inward current is modulated by parameter $E_\text{Na}$
only. The higher the peak voltage, the higher the value of
$E_\text{Na}$. This is consistent with our finding (Figure \ref{fig2}
and Table \ref{table1}), where the LR \cite{Luo-Rudy-1991} model has the highest
$E_\text{Na}$ because the model exhibits the largest action potential amplitude.
Plateau phase is the long phase of the action potential during which
the membrane potential remains depolarised and changes more
slowly. This occurs due to balance between some inward and outward
currents \citep{Santana-2010}. In the archetypal model, $E_{\ast}$ is one of the
parameters that controls the voltage value at the plateau region. The
estimated value is consistent for each model, where $E_{\ast}$ in the
LR \cite{Luo-Rudy-1991} model is the biggest since the LR model has
the largest value of voltage during this phase, while the smallest is
shown by the Noble model.

To fit the archetypal model to isolated cardiac ventricular cell
data we use a pacing rate of 0.3Hz ($B=3333$ ms) at both healthy and
heart-failure conditions as this is the basic cycle length employed in 
\citep{McIntosh-2000}. Figure \ref{fig2}(d) shows the action potential 
morphologies after the fitting process and the new estimated parameter
values are shown in last two columns of Table \ref{table1}. Overall,
the archetypal model exhibits good correspondence with the
targeted data, with minor discrepancies in the plateau region. The average
relative error between the action potentials is relatively small as
seen in Table \ref{table1}(b). 
Figure \ref{fig3}(c) shows that after fitting the archetypal model to
heart-failure data the parameter values most strongly affected are
those related to the $n$ gating variable (the slow-gating potassium
channels), namely $k_1$, $k_2$, $k_3$, $r$ and $g_{22}$. These
parameters need to be adjusted in order to compensate the large
$\text{APD}_{90}$ exhibited in the heart-failure group, which is
commonly reported due to down-regulation of potassium current
\citep{Beuckelmann-Erdmann-1993, Akar-Rosenbaum-2003}. Other
archetypal parameter values like $E_\text{Na}$ and $E_1$ are similar
in both heart-failure and healthy cells since the action potentials
have identical amplitude and resting membrane potential.

\subsection{Tests of APD restitution and computational speedup}

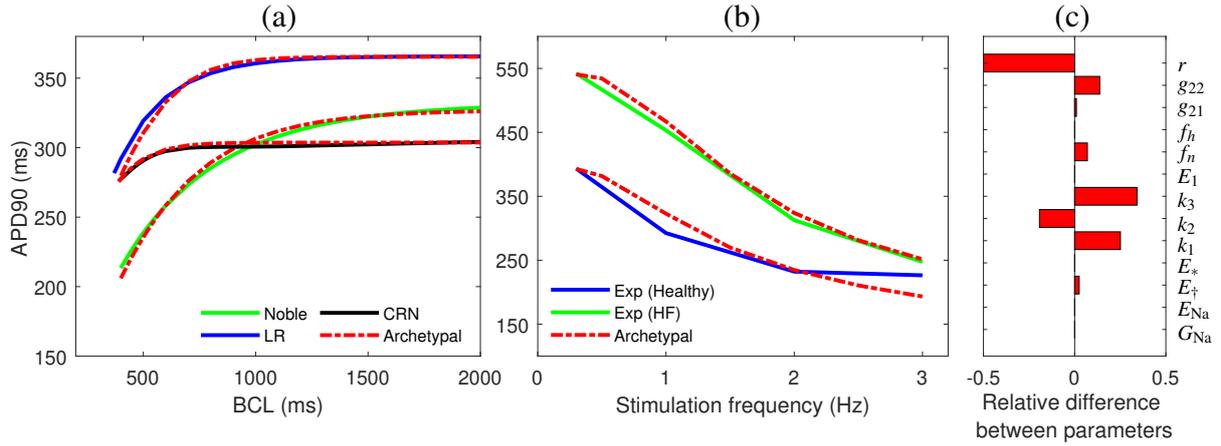
\begin{figure}[t]
\centering
\begin{overpic}[width=\textwidth]{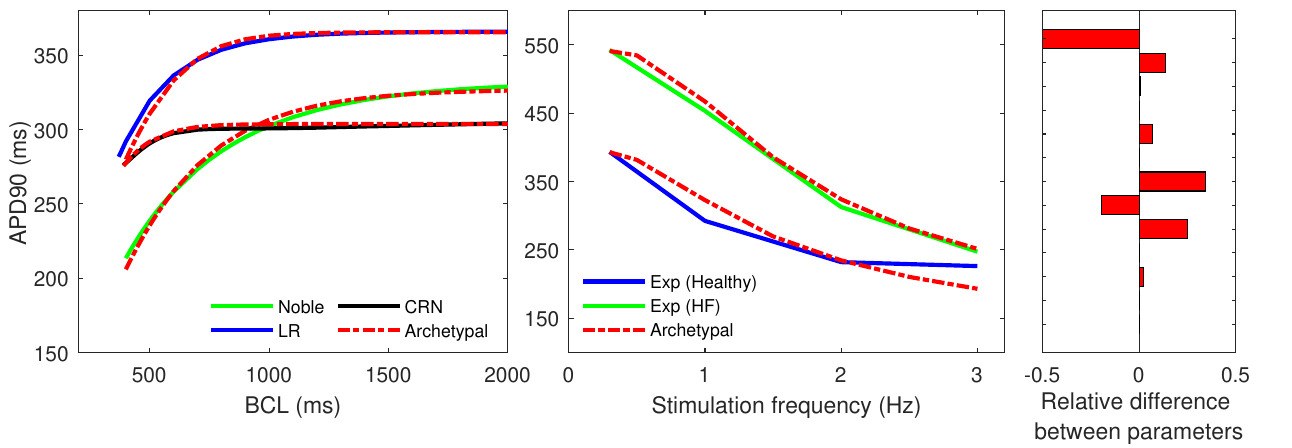} 
\put (97,158) {{(a)}}
\put (270,158) {{(b)}}
\put (394,158) {{(c)}}
\put (440,141) {\scriptsize{$r$}}
\put (440,134) {\scriptsize{$g_{22}$}}
\put (440,125.5) {\scriptsize{$g_{21}$}}
\put (440,116) {\scriptsize{$f_{h}$}}
\put (440,108) {\scriptsize{$f_{n}$}}
\put (440,99) {\scriptsize{$E_{1}$}}
\put (440,90) {\scriptsize{$k_{3}$}}
\put (440,81) {\scriptsize{$k_{2}$}}
\put (440,73) {\scriptsize{$k_{1}$}}
\put (440,65) {\scriptsize{$E_{\ast}$}}
\put (440,57) {\scriptsize{$E_{\dagger}$}}
\put (440,49) {\scriptsize{$E_\text{Na}$}}
\put (440,40) {\scriptsize{$G_\text{Na}$}}
\end{overpic}
\caption{(a,b) APD restitution curves for the archetypal model
  (dash-dotted line)  compared to APD restitution curves of target
  models/data (solid lines coloured as specified in panel legends). (c) Relative
  difference  between the set of archetypal model parameter values
  corresponding to experimental measurements in cells from healthy and
  heart-failure groups.
}
\label{fig3}
\end{figure}

In order to test the validity of our parameter estimation results
beyond the conditions at which they are obtained, we compare
the action potential duration (APD) restitution curves for each of the
target models and the experimental data to corresponding curves
computed  using the parameter estimates reported in Table
\ref{table1}. Figure  \ref{fig3}(a) demonstrates excellent agreement
between the restitution curves of the authentic target models and the
fitted archetypal models, while  Figure  \ref{fig3}(b) shows similarly
good agreement with the restitution curves of the cells from healthy
and heart-failure groups.  In particular, at a given stimulation
frequency failing myocytes exhibit larger 
$\text{APD}_{90}$ than healthy myocytes. At high stimulation frequency
the $\text{APD}_{90}$ values for both groups show a less pronounced
difference. The archetypal model is slightly better able to reproduce
the accurate APD 
restitution curve for the HF myocyte, compared to healthy myocyte. For 
healthy myocyte, the discrepancy occurs at several stimulation
frequencies, and it gets pronounced at stimulation frequency larger
than 2 Hz, where the archetypal model produces smaller $\text{APD}_{90}$
than the experimental data from healthy myocytes.

Because of the practical value of APD restitution curves, it is
important that the fitted archetypal model is able to reproduce the
restitution behaviour of the target models and data. 
These curves describe the dependence of the APD on the
duration of the preceding diastolic interval (DI).
Nolasco and Dahlen \cite{Nolasco1968} noted that in a single-cell
setting and with a fixed period of excitation, a slope of the APD(DI)
curve greater than one indicates instability of the train of action
potentials. For this reason, the restitution curves are considered an
important tool in understanding instabilities of excitation waves
leading to onset of cardiac arrhythmias \cite{Tse2016} and is routinely measured
experimentally, e.g. in the experimental work \cite{McIntosh-2000}
that we compare with.

In order to quantify the computational speedup gained by using the
archetypal model with fitted parameter values in comparison with
authentic target models we measured the time taken using each model to compute
numerically a train of 1000 action potentials at a fixed value of the basic cycle
length $B=1000$ ms. With the numerical methods described above in the
paper, the archetypal model takes approximately about 180 sec to complete
the numerical simulation with any set of parameter values. The LR
\cite{Luo-Rudy-1991} and CRN \cite{Courtemanche-1998} model require
1246 sec and
1634 sec, respectively for the same simulation. Thus we conclude that
using the archetypal model is 6 times faster than using the Luo-Rudy model
\cite{Luo-Rudy-1991} and 9 times faster than the using  Courtemanche
et al.~model \cite{Courtemanche-1998}.
\rev{While the comparison was demonstrated at the single-cell level,
  we expect the improvement in the computational speed-up would also
  be seen in simulations of whole-heart or 3D tissue with realistic geometries.} The numerical
speedup advantage of using the archetypal model is likely even more
pronounced in case of comparison with contemporary models more
detailed than the LR \cite{Luo-Rudy-1991} and the CRN \cite{Courtemanche-1998} models. We also recall that the
archetypal has closed form solutions that can be evaluated directly
irrespective of parameter values used and this can be exploited to
further reduce computational expenses or even eliminate the need of
computation entirely.

\section{\rev{Discussion}}

\subsection{\rev{Summary}}

Contemporary mathematical models of single-cell cardiac electrical
excitation have become immensely detailed. Along with increasing
physiological realism such model complexity leads to parameter
value uncertainty, high computational cost and barriers to mechanistic
understanding. There is thus a need for conceptually and
mathematically simple but physiologically accurate reduced models of
the cardiac action potential. A single-cell cardiac excitation model
that replicates the phase-space geometry of detailed cardiac models 
but is much simpler in both functional form and number of free
parameters was derived in  \cite{Biktashev-2008} and applied to a
number of idealised problems
\cite{Simitev2011,Bezekci2015,Mortensen2021,Mortensen2021b}. 
In order to render the archetypal model of \cite{Biktashev-2008}
also practically applicable to the description of physiological
measurements and to whole-heart and tissue numerical simulations, in
this study we report a robust method for estimation of the parameter
values of the model so as to approximate the action potential
biomarkers of contemporary detailed ionic models as well as
experimental data from direct wet-lab cell measurements.
The parameter estimation procedure relies on the well-known and popular Nelder-Mead 
method \cite{Nelder-Mead-1965} for minimisation of multi-variable
functions by direct simplex search. Here, the optimal parameter
values of the archetype are determined by minimising the residual difference
between the morphologies of single-stimulus action potentials of the
archetypal model and the target model/data.
The procedure is then applied to 5 test cases, namely (a) to the authentic Noble
model \cite{Noble1962}, the precursor to all detailed cardiac ionic current
models; (b) to the Luo-Rudy ventricular cell model \cite{Luo-Rudy-1991}, the original second generation
model; (c) to the Courtemanche, Ramirez and Natel atrial cell model
\cite{Courtemanche-1998}, a popular modern cardiac system, as well as
(d,e) to wet lab experimental measurements of rabbit ventricular cells from
both healthy and heart-failure samples \cite{McIntosh-2000}.
In all cases, sets of values of the archetypal model parameters have
been found so that the morphologies of stable single-stimulus action
potential target transients are well reproduced.
As this is an optimisation problem in high-dimensional parameter space
that may have several local minima, it is difficult to ascertain if a
given solution of the minimisation procedure using Nelder-Mead method
is the best possible one. To further test and 
validate the method, action potential duration restitution curves
are also computed and compared to those of the target models and data,
again with excellent agreement.
We conclude that compared to more
sophisticated parameter estimation methods for cardiac models such
as maximum-likelihood estimation \citep{Milescu-Sachs-2005},
principal-axis fitting \citep{Vandenberg-Bezanilla-1991}, genetic
algorithms \citep{Syed-2005, Kaur2014}, as well as the widely
practised empirical ``hand-tuning'' of free parameters
\citep{Baranauskas-Martina-2006,Mickus-Spruston-1999},  
our rather straightforward approach provides comparable quality of
approximation and performs remarkably well. This is likely due to the generic
structure and the small number of parameters of the archetypal model
we consider.
An open-source Matlab(TM) implementation of the models and methods is
made permanently available at \cite{Aziz2021} and can be used by
the readers to fit the archetypal model to models and data of their own choice. 

\rev{Many processes that occur in excitable cells, including cardiac cells
are still not fully understood. None of the detailed models are
themselves ultimate, rather they are continually improved and in some
cases discarded in the light of new experimental measurements. The approach of parameter adjustment used
in the present work, is a way to accurately model cellular electrical
excitation even if fine details of cell physiology are not included.}

\subsection{\rev{Perspectives for future work}}

\rev{A major advantage of the archetypal model \eqref{model} is that it has three features that, to our
  knowledge, were not available in combination for any other model prior to this work.
{
    \setlength{\parskip}{-5pt}
    \setlength{\parsep}{-5pt}
\begin{enumerate}
    \setlength{\itemsep}{0pt}
    \setlength{\parskip}{0pt}
    \setlength{\parsep}{0pt}
  \item[(a)] Model \eqref{model}  admits both asymptotic and closed-form analytical
    solutions, see \cite{Biktashev-2008} and the extended discussion
    in the Introduction.
  \item[(b)] Model \eqref{model} captures essential cardiac excitability
    characteristics such as slow repolarization, slow subthreshold
    response, fast accommodation, variable peak voltage, and front
    dissipation that other ad hoc simplifications do not, see \cite{Biktashev-2008}.
  \item[(c)] With the results of the present work, model \eqref{model}
    can now be fitted to reproduce accurately the
    electrophysiological responses of a variety cardiac cell types.
\end{enumerate}
There is a large body of already developed theory for conceptual
understanding of the dynamics of  nonlinear wave processes in cardiac
tissue that underlie arrhythmias, fibrillation and defibrillation \cite{Qu2014}. The
unique features of our model open the way to apply this theory to
realistic experimental and clinical situations with a dramatic
increase of quantitative accuracy. 
Particular examples include: applying known mathematical conditions
of propagation block in terms of fitted myocyte parameters, making
realistic analytical estimates of the vulnerability to 
extrastimuli, realistic prediction of the frequency and stability of
functional re-entrant circuits and likelihood of recurrent
fibrillation after a defibrillating shock. Similarly, robust
relationships between controllable cell parameters and outcome of
experiments can now be obtained that will have the
potential to allow more confident planning of experiments and
facilitate development and improvement of antiarrhythmic
strategies. These applications are planned for future work.}}

\rev{We wish to comment, in particular, on the perspectives that our works
opens for the development of novel efficient numerical methods for
excitation propagation and tissue simulations. The archetypal model
\eqref{model} allows computational speed-up to be achieved in two
essentially different ways.
{
    \setlength{\parskip}{-6pt}
    \setlength{\parsep}{-6pt}
\begin{enumerate}
    \setlength{\itemsep}{0pt}
    \setlength{\parskip}{0pt}
    \setlength{\parsep}{0pt}
  \item[(a)] Firstly, a straightforward replacement of large detailed
ion current models by model \eqref{model} with appropriately fitted
parameters will lead to a several-fold speed-up as measured above for
the LR and the CRN models. This is already a significant improvement
as a speed-up of 6 to 9 times is comparable to the speed-up of using
e.g. lookup  tables \cite{Clayton-2011}.
  \item[(b)] Secondly, more important benefits may be achieved by
employing the asymptotic structure readily encoded in \eqref{model} to split
the model to a fast-time subsystem describing only the front of the
action potential coupled to a slow-time subsystem describing its plateau
and recovery phases.
\end{enumerate}
To explain the importance of (b), we note that physiological cell
ionic models are stiff because the 
dynamics of the action potential front is orders of magnitude faster
than the dynamics during the plateau and recovery. This requires 
very small time and spacial discretisation steps to be used in
numerical schemes to adequately resolve propagating
action potentials.}
After asymptotic splitting of the archetypal model \eqref{model}, a
numerical scheme can be used that will require a fine resolution only
for solution of the fast-time front subsystem and allow a much
coarser resolution to be used for solution of the non-stiff plateau and recovery
subequations, e.g. the heterogeneous multiscale method \cite{Weinan2007}. 
However, hybrid asymptotic-numerical methods are not well developed in higher
dimensions, which are needed for calculation of activation
sequences. The equation 
of motion for the front is a partial differential equation of motion of a line (in 2D)
or surface (in 3D). One immediate difference to the 1D case is that propagation of the front 
no longer depends only on the pre-front voltage and slow variables,
but also on its own spatial configuration. Fortunately, unless the shape of
the front deviates  very strongly from plain, the effect of its shape
can be taken into account via its mean curvature that can be
easily incorporated \cite{Simitev2006}. This approach can be used to describe
normal activation sequences in the heart, when the graph of the front
solution in the space-time is a manifold without internal
discontinuities. More serious challenges occur if there are propagation
blocks and/or wave breaks, which introduce discontinuities of the front
manifold in space-time. In such cases, a separate asymptotic
description for the codimension-two areas, the wave break trajectories
and the propagation block loci, are needed; obtaining such asymptotic
description is another important direction for further research.
These issues will be much easier to tackle using our 
simple (and now accurate) archetypal model rather than complex physiologically
detailed cell ionic models.} 

\section{Acknowledgements}

This work was supported by the UK Engineering and Physical Sciences
Research Council [EPSRC grant numbers EP/N014642/1,   EP/S030875/1,
  EP/T017899/1]. M.H.N. Aziz was funded by the Ministry of Higher
Education Malaysia and Universiti Malaya under a SLAB scholarship. 

\section{References}

\renewcommand{\bibsection}{}
\setlength{\bibsep}{0.5mm}

\authorsNames{Muhamad H.N. Aziz,, Radostin D. Simitev}

\authorsEmails{\BIOauthorEmail{hifz$\_$din@um.edu.my},%
\BIOauthorEmail{},%
\BIOauthorEmail{Radostin.Simitev@glasgow.ac.uk}}

\authorsPhotos{photoDin05,nophoto,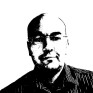}

\authorsBios{{Aziz graduated from Universiti Malaya, Malaysia
    with a Bachelor in Mathematics in 2014, followed by a Master in
    Applied Mathematics in 2015 from the University of Glasgow, United
    Kingdom. He recently received his Ph.D in 2021 from the same
    university in the area of mathematical biology. He currently holds
    a lecturer position at Universiti Malaya. His research interest
    is in the modelling of cardiac electrophysiology, particularly in
    investigating the cellular heterogeneity and their responses to
    cardiac antiarrhythmic drugs.},%
  {},%
  {\rev{Simitev (orcid.org/0000-0002-2207-5789) is a Professor of
      Applied Mathematics at the University of Glasgow.} He works in
    the area of magnetohydrodynamics and convective dynamo theory,
    where he has made contributions to modelling of geomagnetic
    polarity reversals, solar cycles and bistability of turbulent
    dynamos. Simitev also works in cardiac excitability and
    arrhythmogenesis where he has used asymptotic methods to
    investigate absolute refractoriness in atrial tissues, conduction
    velocity restitution and initiation of excitable waves.} 
}

\biographyAuthors

\licen


\end{document}